\documentclass[twocolumn,preprintnumbers,amsmath,amssymb,prd]{revtex4}
\usepackage{graphicx}
\usepackage{multibox}

\def\bbox{{\,\lower0.9pt\vbox{\hrule \hbox{\vrule height 0.2 cm\hskip
0.2 cm \vrule height 0.2 cm}\hrule}\,}}
\newcommand{\dsl}{\pa \kern-0.5em /}

\def\ben{\begin{equation}}
\def\een{\end{equation}}
\def\bena{\begin{eqnarray}}
\def\eena{\end{eqnarray}}

\newcommand{\be}{\begin{equation}}
\newcommand{\ee}{\end{equation}}
\newcommand{\bea}{\begin{eqnarray}}
\newcommand{\eea}{\end{eqnarray}}
\newcommand{\beq}{\begin{eqnarray}}
\newcommand{\eeq}{\end{eqnarray}}

\def\6{\partial}

\def\d{\delta}

\def\pa{\partial}
\def\B{\begin{equation}}
\def\E{\end{equation}}

\def\today{\ifcase\month\or  January\or February\or March\or April\or
May\or June\or  July\or August\or September\or October\or
November\or December\fi \space\number\day, \number\year}
\input amssym.def
\input amssym.tex

\begin{document}
\preprint{CECS-PHY-07/05} \preprint{ULB-TH/07-13}
\title{A Monopole Near a Black Hole}
\author{Claudio Bunster}
\affiliation{Centro de Estudios Cient\'{\i}ficos (CECS), Valdivia,
Chile.}
\author{Marc Henneaux}
\affiliation{Physique Th\'eorique et Math\'ematique, Universit\'e
Libre de Bruxelles \& International Solvay Institutes, ULB Campus
Plaine C.P. 231, B--1050 Bruxelles, Belgium, and \\Centro de
Estudios Cient\'{\i}ficos (CECS), Valdivia, Chile.}

\begin{abstract}

A striking property of an electric charge near a magnetic pole is
that the system possesses angular momentum even when both the
electric and the magnetic charges are at rest.  The angular
momentum is proportional to the product of the charges and
independent of their distance. We analyze the effect of bringing
in gravitation into this remarkable system. To this end, we study
an electric charge held at rest outside a magnetically charged
black hole.  We find that even if the electric charge is treated
as a perturbation on a spherically symmetric magnetic
Reissner-Nordstrom hole, the geometry at large distances is that
of a magnetic Kerr-Newman black hole.  When the charge approaches
the horizon and crosses it, the exterior geometry becomes that of
a Kerr-Newman hole with electric and magnetic charges and with
total angular momentum given by the standard value for a charged
monopole pair. Thus, in accordance with the ``no-hair theorem",
once the charge is captured by the black hole, the angular
momentum associated with the charge monopole system, looses all
traces of its exotic origin and it is perceived from the outside
as common rotation. It is argued that a similar analysis performed
on Taub-NUT space should give the same result, namely, if one
holds an ordinary mass outside of the horizon of a Taub-NUT space
with only magnetic mass, the system, as seen from large distances,
is endowed with an angular momentum proportional to the product of
the two kinds of masses.  When the ordinary, electric, mass
reaches the horizon, the exterior metric becomes that of a
rotating Taub-NUT space. This rotating space, ``Kerr-Taub-NUT
geometry", is a solution of the vacuum Einstein equations
different from ordinary Taub-NUT space (Taub-NUT space, does not
possess angular momentum in spite of having both electric and
magnetic mass).  In the process of performing the calculation, an
economic way to compute the angular momentum is provided, which
also clarifies possible confusion about Dirac strings and the
like.

\end{abstract}

\maketitle

%%%%%%%%%%%%%%%%%%%%%%%%%%%%%%%%%%%%%%%%%%%%%%%%%%%%%%%%%%%%%%%%%%%%%%%%%%%
\section{Introduction}
\setcounter{equation}{0}
%%%%%%%%%%%%%%%%%%%%%%%%%%%%%%%%%%%%%%%%%%%%%%%%%%%%%%%%%%%%%%%%%%%%%%%%%%%

Magnetic poles and black holes are remarkable objects.  To some
extent they have had similar histories.

The black hole emerged as a solution of the Einstein equations which
was at first regarded as unphysical because of its singular nature.
However, further study for many years by many researchers
demonstrated that black holes were physically relevant as the
endpoint of gravitational collapse (see e.g. \cite{Misner:1974qy}).
But even then, collapsed objects were thought to be scarce and the
observational search for them began. Nowadays it is commonly
accepted that enormous black holes exist at the centers of galaxies,
including our own \cite{BigBlackHoles}. Furthermore, those black
holes may be actually responsible for the very existence of the
galaxies themselves and therefore, for the presence of structure in
the universe.

It is no minor feat that in less than seventy years
\cite{Opp-Solvay} the black hole has risen from the status of an
unphysical exotic solution of the Einstein equations to being an
observed astrophysical object responsible for structure in the
universe!

The magnetic pole was first introduced as an appealing
modification of Maxwell's equations which, without it, are not
fully symmetric under duality rotations of the electric and
magnetic fields \cite{Dirac:1931kp,Dirac:1948um}.  The
introduction of the magnetic pole had a spectacular implication,
namely, that the mere existence of a single magnetic pole in the
universe would imply that all electric charges should be integer
multiples of a basic quantum inversely proportional to the
magnetic charge of the pole. This provided the first possible
theoretical basis for an hitherto totally unexplained key feature
of the universe, the quantization of electric charge.

However, the introduction of magnetic poles was a modification
which, although attractive for aesthetic reasons was not implied
by Maxwell's equations themselves which had been amply validated
by experiment. But then, help came from a different avenue of
inquiry, namely the attempt to find a theory that incorporated the
charge independence of nuclear forces, which gave rise to the
Yang-Mills fields. Magnetic poles were shown to arise as solutions
of the field equations of an $SU(2)$ Yang-Mills theory coupled to
a scalar field multiplet.  The solution was regular and the
electric and magnetic charges obeyed the Dirac quantization
condition \cite{'tHooft:1974qc,Polyakov:1974ek}. At that point,
the focus of Yang-Mills theory had already shifted from the charge
independence of the nuclear forces (gauge theory of isospin) to
the interaction of subnuclear matter, and Grand Unified Theories.
These GUTs predict the existence of monopoles (see e.g. Chapter 23
of \cite{Weinberg}).

So it would appear founded to say that the magnetic pole has
gained full theoretical respectability and that it is now not an
option but a consequence of accepted microphysical theory.
However, this success story is not at the same level as that of
the black hole.  Indeed, magnetic poles have not been observed
and, more importantly, we lack a distinct fundamental role for
them in our present view of the universe.

Nevertheless, if the history of the black hole is to teach us a
lesson, it is that it might not be totally out of the question to
think that this simple fundamental object has not yet found its
proper central place in physics but it should at some point do so.

With the above motivation in mind, we have undertaken the study of
the simplest problem where these two remarkable objects, the black
hole and the magnetic pole, interact.

The plan of the paper is as follows.  We first recall (section
{\bf \ref{Flat}}) the striking property of an electric charge near
a magnetic pole in flat space, which is that the system possesses
angular momentum even when both the electric and the magnetic
charges are at rest. The angular momentum is proportional to the
product of the charges. In this process we develop an economic way
to compute the angular momentum, which also clarifies possible
confusion about Dirac strings and the like.

Next, in section {\bf \ref{ElectricBH}}, we study an electric
charge held at rest outside a magnetically charged black hole.
This situation is equivalent, by electromagnetic duality, to the
case of a magnetic pole placed at rest in the background of an
electric black hole announced in the title of the paper.  We find
that even if the electric charge is treated as a perturbation on a
spherically symmetric magnetic Reissner-Nordstrom hole, the
geometry at large distances is that of a magnetic Kerr-Newman
black hole.  When the charge approaches the horizon and crosses
it, the exterior geometry becomes that of a Kerr-Newman hole with
electric and magnetic charges and with total angular momentum
given by the standard value for a charged monopole pair. Thus, in
accordance with the ``no-hair theorem", once the charge is
captured by the black hole, the angular momentum associated with
the charge monopole system, looses all traces of its exotic origin
and it is perceived from the outside as common rotation.

Section {\bf \ref{higherspin}} is devoted to arguing that a
similar analysis performed on Taub-NUT space would give the same
result, namely, if one holds an ordinary mass outside of the
horizon of a Taub-NUT space with only magnetic mass, the system,
as seen from large distances, is endowed with an angular momentum
proportional to the product of the two kinds of masses.  When the
ordinary, electric, mass reaches the horizon, the exterior metric
becomes that of a rotating Taub-NUT space. This rotating space,
``Kerr-Taub-NUT geometry", is a solution of the vacuum Einstein
equations different from ordinary Taub-NUT space (Taub-NUT space
does not possess angular momentum in spite of having both electric
and magnetic mass).

Finally, section {\bf \ref{conclusions}} is devoted to brief
concluding remarks.  Among them we observe that through successive
captures of electric and magnetic poles of both signs, a
Schwarzschild black hole can become a (neutral!) Kerr hole.  We
then indulge in the speculation that monopoles might account for
some of the rotation of the black holes in the universe.

\section{Monopole angular momentum revisited}
\setcounter{equation}{0} \label{Flat}

We consider a magnetic pole of strength $g$ at the origin of
coordinates in flat space and an electric charge $q$ located at a
distance $c$ above the magnetic pole on the $z$-axis.  The
magnetic and electric fields are \be \overrightarrow{B} = \frac{g
\, \vec{r}}{
 r^3}\,, \qquad \overrightarrow{E} = \frac{q \,
(\vec{r}-\vec{c}) }{ |\vec{r}-\vec{c}|^3}\, \label{BandE}\ee with
$\vec{c} = c \, \hat{e}_3$.  They obey the Gauss law equations \be
\overrightarrow{\nabla} \cdot \overrightarrow{B} = 4 \pi \, g \,
\delta^{(3)} (\vec{r}) \,, \qquad \overrightarrow{\nabla} \cdot
\overrightarrow{E} = 4 \pi \, q \, \delta^{(3)} (\vec{r} -
\vec{c})\, . \ee We will use the standard ``electric picture" and
introduce a vector potential for the magnetic field which only has
a non vanishing azimuthal component, \be \label{vecpot}A = g \, (k
- \cos\theta)\, d\phi \, \ee as well as the magnetic and electric
field densities \be \mathcal{B}^i = \sqrt{g}\, B^i \,, \qquad
\mathcal{E}^i = \sqrt{g}\, E^i \, .\ee The formula \be
\mathcal{B}^i = \varepsilon^{ijk} \partial_j A_k \label{BfromA}
\ee reproduces the field given in (\ref{BandE}). However, the
potential (\ref{vecpot}) is well-defined only away from the
$z$-axis for a general value of $k$.  The singularity of the
potential on the $z$-axis may be pictured in physical terms as a
concentrated flux coming in along the positive and negative
$z$-axes with strengths that add up to $g$.  This flux re-emerges
then radially from the origin to give, away from the $z$-axis, the
field (\ref{BandE}). To compensate for this singular flux, one
normally brings in an additional entity, the Dirac string,  which
cancels the flux and therefore the right side of (\ref{BfromA})
acquires an additional contribution in order to be valid also on
the $z$-axis. It is important to emphasize that the Dirac string
is not the singularity of $A_\phi$ but, rather, the additional
object which is brought in in order to cancel it.  As a
consequence, if one is only interested in the $B$-field given in
(\ref{BandE}), as we will be in the present paper, one may just
take (\ref{BfromA}) as is and simply extrapolate continuously its
value to the $z$-axis.

Thus, in computing the angular momentum stored in the field \be
\label{belin} {\vec J} = - \frac{1}{4 \pi}\int_V \vec{r} \times
(\vec{E} \times \vec{B})\, d^3 x \ee we may substitute
(\ref{BfromA}) for $B$ and obtain the correct answer as we shall
proceed to do.

{}For symmetry reasons, the only non vanishing component of the
angular momentum ${\vec J} $ will be along the $z$-axis.
Customarily, one tackles the integral directly in Cartesian
coordinates.  However, it will be simpler, and useful to us
further below, to work in spherical coordinates recalling that the
$z$-component of the angular momentum is simply the azimuthal
component of the linear momentum, whose density is the Poynting
vector \be - \frac{1}{4 \pi} F_{\phi j} \mathcal{E}^j \, . \ee

Therefore, the only non vanishing component of (\ref{belin}) reads
\be J_z = - \frac{1}{4 \pi} \int dr\, d\theta \, d\phi F_{\phi j}
\mathcal{E}^j \, . \label{Jz}\ee

We evaluate the integral as follows.  First, we note that
$\mathcal{B} $ and $A_\phi$ depend only on $\theta$ while
$\mathcal{E}$ depend only on $r$ and $\theta$. {}Furthermore, from
(\ref{BandE}), the only vanishing components of $\mathcal{E}^i$
are $\mathcal{E}^r$ and $\mathcal{E}^\theta$ so that we can
rewrite (\ref{Jz}) as, \be J_z = \frac{1}{4 \pi}  \int dr\,
d\theta \, d\phi \,
\partial_\theta A_\phi \, \mathcal{E}^\theta \, . \ee  Now we observe from
(\ref{BandE}) that $\mathcal{E}^\theta$ vanishes at $\theta = 0$
and $\theta = \pi$. Therefore, after integration by parts in
$\theta$, we obtain \be J_z =  - \frac{1}{4 \pi} \int dr\, d\theta
\, d\phi \,
 A_\phi \, \partial_\theta \mathcal{E}^\theta \, . \label{II10}\ee  Next we
 rewrite (\ref{II10}) by introducing the explicit value (\ref{vecpot}) for
 $A_\phi$ and repeatedly using Gauss'law for the electric field which, written in spherical
 coordinates, reads \be \partial_r \mathcal{E}^r + \partial_\theta \mathcal{E}^\theta =
 4 \pi \, q \, \delta(r-c) \, \delta^{(2)}(\theta, \phi) \, \ee where
 $\delta^{(2)}(\theta, \phi)$ is the $\d$-function density defined
 on the sphere with support at the North Pole, that is $\int
 f(\theta, \phi) \delta^{(2)}(\theta, \phi) = f(\theta=0)$.
 We obtain for the angular momentum contained in a region
 of space bounded by the two spheres of radii $r_1$ and $r_2$
\be J_z = \int_{r_1}^{r_2} dr j(r) \ee with \be j = \frac{d}{dr}
\left[   gq  H(r-c) -\frac{g}{ 2 }  \int_0^\pi d \theta \cos
\theta \mathcal{E}^r(r, \theta) \right] \label{dens1}\ee where
$H(r-c)$ is the Heaviside step function ($H=0$ for $r<c$, $H=1$
for $r>c$). Remark that the constant $k$ drops out of the final
answer as it should be the case since the magnetic field does not
depend on it.

Expression (\ref{dens1}) for the effective radial density $j(r)$
of $J_z$ has the remarkable property of being the derivative
(``divergence") of a local function of $r$, which means that one
can write the angular momentum contained between the two spheres
as the difference between two ``surface integrals" (``fluxes"),
namely, \be J_z = \Phi(r_2) - \Phi(r_1) \label{JPhi}\ee with \be
\Phi(r) =  gq H(r-c) - \frac{g}{2}\int_0^\pi d \theta \cos \theta
\mathcal{E}^r(r, \theta) \, . \ee  Note in particular that \be
\Phi(0) = 0 \, . \label{Flux0}\ee

Eq. (\ref{Flux0}) follows from : (i) the Heaviside function
vanishes because $c>0$, (ii) near the origin, $ \mathcal{E}^r = q
\frac{r^2}{c^2} \sin \theta \, \cos \theta $, which vanishes at
$r=0$. It states that there is no $\d$-function source for the
angular momentum on the magnetic pole. By symmetry, there is no
$\d$-function source for the angular momentum at the electric pole
either.

As a consequence of (\ref{JPhi}) and (\ref{Flux0}), the angular
momentum within a sphere of radius $r$ is \be J_z = \Phi(r) \ee
and, for the total angular momentum, we recover the standard
result \be J_z = \Phi(\infty) =  q g\, . \label{Jztotal}\ee This
follows from : (i) the Heaviside function is equal to one because
$c< \infty$, (ii) near infinity, $ \mathcal{E}^r = q \sin \theta$,
and this expression, once multiplied by $\cos \theta$ and
integrated over $\theta$, yields zero.  As it is well known, if
one demands that $J_z$ given by (\ref{Jztotal}) should be
quantized to half-integer values, one obtains the Dirac
quantization condition \be \frac{2 q g}{\hbar} \in \mathbb{Z} \ee

\section{Electric charge near a magnetic black hole}
\setcounter{equation}{0} \label{ElectricBH}

We now consider the case of an electric charge $q$ placed at rest
in the background of a black hole with magnetic charge $g$.  Just
as before, we place the electric charge on the positive $z$-axis,
at $r=c$. This situation is equivalent, by electromagnetic
duality, to the case of a magnetic pole placed at rest in the
background of an electric black hole announced in the title of the
paper and permits direct contact with the preceding discussion in
flat space.

The unperturbed geometry is described by the Reissner-Nordstrom
metric, \beq ds^2 &=& - \left(1 - \frac{2M}{r} + \frac{g^2}{r^2}
\right) dt^2 + \left(1 - \frac{2M}{r} + \frac{g^2}{r^2}
\right)^{-1} dr^2 \nonumber \\ && + r^2 (d \theta ^2 + \sin ^2
\theta d \phi^2) \label{RN}\eeq and the unperturbed
electromagnetic field is purely magnetic and given by the
expression, \be F = g \sin \theta \, d\theta \wedge d \phi \, ,
\label{FieldStr}\ee just as in the flat space case.

We first determine, to first order in $q$, the components of the
perturbed fields relevant to the computation of the angular
momentum. To that effect, we observe that the perturbation of the
electromagnetic field is purely electric, and that, just as
before, the electric field $\mathcal{E}^i$, whose divergence gives
a delta function at the location of the charge, has no azimuthal
component. {}Furthermore, $\mathcal{E}^\theta$ and $\mathcal{E}^r$
depend only on $r$ and $\theta$.

In general relativity, the total angular momentum is given by a
surface integral at infinity because it is a global conserved
charge associated with a gauged symmetry. The surface integral is
determined by the requirement that the corresponding generator
should have well-defined functional derivatives
\cite{Regge:1974zd}.   In the case at hand (rotations about the
$z$-axis), the generator can be taken to be \be G= \int d^3x
\xi^\phi \mathcal{H}_\phi  + J_z \, .\ee Here $\xi^\phi$ is an
arbitrary function of the spatial coordinates (``surface
deformations", ``gauged rotations") which tends to unity at
infinity, \be \xi^\phi \rightarrow 1 \, , \;  \; \; \; r
\rightarrow \infty \, .\ee  The Hamiltonian generator
$\mathcal{H}_\phi$ is given by \be \mathcal{H}_\phi = -2 \pi^{\;
\; k}_{\phi \;  \; \;  \vert k} - \frac{1}{4 \pi} F_{\phi k}
\mathcal{E}^k \, .\ee Here, $\pi^{ij}$ is the canonical conjugate
to the spatial metric $g_{ij}$, $\mathcal{E}^k$ is the electric
field density which is proportional to the canonical conjugate
$\pi^k$ of the potential $A_k$ ($\pi^k = \frac{1}{4
\pi}\mathcal{E}^k$), and the vertical bar denotes covariant
differentiation in $g_{ij}$.

The surface integral $J_z$ is determined by the demand that the
variation of the generator $G$ should be given by the volume
integral of a local function containing no derivatives of the
variations of the dynamical variables.  Thus in practice $J_z$ is
constructed so as to compensate for the surface integrals at
infinity which arise upon integration by parts in the volume piece
of $\delta G$. In order to implement this procedure, it is
necessary to give boundary conditions at infinity for all the
fields.  These boundary conditions include definite parity
conditions (behaviour under $\theta \rightarrow \pi - \theta$,
$\phi \rightarrow  \phi + \pi$) \cite{Regge:1974zd}. In particular
the vector potential should be odd to leading order.  This means
that in the Hamiltonian treatment one must take the arbitrary
constant $k$ appearing in (\ref{vecpot}) equal to zero. Therefore,
the vector potential for the field strength (\ref{FieldStr}) will
be taken to be \be \label{vecpot2}A = - g \cos\theta \, d\phi \,
.\ee

To determine $J_z$ in the case at hand, it is sufficient to write
the Hamiltonian generator $\mathcal{H}_\phi$ taking $g_{ij}$ to be
the spatial metric of the background (\ref{RN}) and allowing for a
perturbation $\pi^{\phi j}(r,\theta)$, of the background
(\ref{RN}), which has zero $\pi^{ij}$.  To begin the analysis, we
write $\mathcal{H}_\phi$ explicitly.  The calculation is quite
simple because the symmetrized combinations of the Christoffel
symbols which appear vanish due to the simple form of the
Reissner-Nordstrom metric.  One finds \be \mathcal{H}_\phi = - 2
\pi^{\; \; k}_{\phi \; \;  , k} - \frac{1}{4 \pi} F_{\phi k}
\mathcal{E}^k \, .\ee The variation of $G$ then gives \be \d G =
\hbox{Volume integral } - 2 \int_{S^2_\infty}   \delta \pi^{\; \;
r}_{\phi}  d \theta d \phi \ee from which we conclude that \be J_z
= 2 \int_{S^2_\infty} \pi^{\; \; r}_{\phi}  d \theta d \phi \,
.\label{JzSurface}\ee

In order to evaluate the surface integral (\ref{JzSurface}), we
first integrate the constraint equation \be \mathcal{H}_\phi = 0
\ee over the 2-sphere and from an arbitrary fixed value of $r$ to
infinity. {}For the electromagnetic contribution, we can take over
the results of Section \ref{Flat} and we obtain \be J_z =
\Phi(\infty) - \Phi(r_+) + 2 \int_{S^2(r_+)} \pi^{\; r}_\phi d
\theta d \phi \label{integrals forJz}.\ee

Now, we observe that due to the conservation of angular momentum,
$J_z$ should be independent of the radial coordinate $c$ of the
electric charge.  Indeed, one can imagine displacing the electric
charge in the radial direction from one location to another.  This
could be accomplished, for example, by letting it fall and then
stopping it, or, say, by moving it adiabatically holding it with a
rope.  In either case, the force exerted upon the charge will be
radial and therefore would exert no torque around the origin. This
reasoning applies also in flat space and it is in that case yet
another way, besides dimensional analysis, to realize that the
angular momentum of the electric-magnetic pole pair is independent
of the distance between the monopoles.

It is thus sufficient to evaluate the integrals in (\ref{integrals
forJz}) for $c >> r_+$.  This, of course, is the same as letting
$r_+ \rightarrow 0$ keeping $c$ finite.  The space then becomes
flat and the domain of integration for each of the two integrals
at $r=r_+$ becomes a 2-sphere of vanishing radius, which makes
each integral to vanish because the integrand is regular.  Here,
we are using the word ``regular" in the geometrical sense. This
means that a regular density of positive weight vanishes at the
origin when expressed in polar coordinates.   So, we find again
\be J_z = \Phi(\infty) = q g \label{valueofJ}\ee (for any $c$).

We may think of formula (\ref{integrals forJz}) as stating that
the total angular momentum is composed of two parts, the angular
momentum stored in the electromagnetic field \be \Phi(\infty) -
\Phi(r_+), \label{angularField} \ee and the spin of the black
hole, \be 2 \int_{S^2(r_+)} \pi^{\; r}_\phi d \theta d \phi
\label{spin}\ee

Imagine now that the point charge is moved toward the horizon and
crosses it.  Once the charge is inside the horizon, we are faced
with the Einstein-Maxwell equations with both electric and
magnetic charges with two Killing vectors
$\frac{\partial}{\partial t}$, $\frac{\partial}{\partial \phi}$.
By the black hole uniqueness theorem, the exterior solution is
then the Kerr-Newman metric \cite{KN,Misner:1974qy} with electric
and magnetic charges with a corresponding electromagnetic field,
linearized in the electric charge and with a value for the total
angular momentum given by (\ref{valueofJ}). That line element will
be explicitly displayed in (\ref{KerrNewman}) below.

What happens is that when the charge is far away from the horizon,
one has a non rotating black hole and the angular momentum is all
stored in the electromagnetic field outside the horizon. As the
charge is brought in, the angular momentum in the field starts
being continuously transferred to the hole which begins to spin
around faster and faster as the charge gets closer and closer to
$r_+$. Thus, (\ref{angularField}) decreases in magnitude from $q
g$ to the value that it has for Kerr-Newman (with both $q$ and
$g$).  This value is not zero as it would be for
Reissner-Nordstrom, since due to the rotation there is a non zero
component ${\cal E}^\theta$.  At the same time, (\ref{spin})
increases from zero to \be 2 \int_{S^2(r_+)} \pi^{\; r}_\phi d
\theta d \phi = - M a + \frac{2}{3} \frac{g^2 a }{r_+}
\label{spinvalue} \ee  where \be - Ma = q g \, .\ee  The spin of
the hole (\ref{spinvalue}) differs from the total angular momentum
by the residual angular momentum in the Kerr-Newman
electromagnetic field \be \Phi(\infty) - \Phi(r_+) = - \frac{2}{3}
\frac{g^2 a }{r_+} \label{angularFieldvalue}. \ee  In order not to
interrupt the thread of the argument, the derivation of
(\ref{spinvalue}) and (\ref{angularFieldvalue}) is given in the
Appendix.

When the charge reaches the horizon, the transfer has become
complete and the black hole is rotating exactly at the required
rate so that the charge can go in smoothly without giving the hole
a jolt. It is as if a child wants to get on a merry-go-round
without hitting himself when he jumps on it. He must then run so
that he reaches the platform with the same angular velocity as the
merry-go-round. The trick here is of course that gravity does the
job for the child by adjusting the angular velocity of the
merry-go-round so that the child can approach in any way he wishes
(even radially!).

It is important to realize how different the situation is for the
black hole case from the flat space case described in the previous
section.  In flat space when the electric charge approaches the
magnetic pole, the electromagnetic angular momentum density is
changed and tends to pile up near the origin. However, the
integral of that density is unchanged and therefore the total
angular momentum is not transferred from the electromagnetic field
to anything else. The pair $(q,g)$ does not start spinning around
as the charges get closer.  It just stays at rest.  On the other
hand, in the black hole case, the hole acquires an intrinsic spin
which leaves the same imprint on the geometry as the one that
would occur if the hole had been formed by the collapse of a
rotating star. The ``transfer" only exists in the presence of the
gravitational coupling which provides the mechanism for its
occurrence and which is also responsible for the existence of the
black hole to begin with.

To end this section, it should be made clear that, for any
location of the electric charge, one may go to radial distances
well beyond it towards infinity. At those large distances the
metric coincides with the asymptotic form of the Kerr-Newman line
element, which reads explicitly (when linearized in $q$) \beq ds^2
&=& -\frac{\Delta}{r^2} dt^2  + r^2 \sin^2 \theta d \phi^2 \nonumber \\
&& + \frac{2 a \sin^2 \theta}{r^2} \left[- 2 M r + g^2
\right] dt d \phi \nonumber \\
&& + \frac{r^2}{\Delta} dr^2 + r^2 d \theta^2 \label{KerrNewman}
\eeq with \be  \Delta = r^2 - 2 M r +g^2 . \ee  As the charge
moves in and the ``hair" progressively disappears, the
approximation of the actual metric by the Kerr-Newman line element
becomes more and more accurate for all distances until it is exact
when the charge reaches the horizon.

\section{Higher spin poles}
\setcounter{equation}{0} \label{higherspin}

It has been shown recently that one may extend the notion of a
magnetic pole to higher integer spin gauge fields \cite{BCHP}. For
spin $s$, the corresponding ``electric" and ``magnetic" conserved
charges are symmetric tensors of rank $(s-1)$ $P_{\nu_1 \nu_2
\cdots \nu_{s-1}}$ and $Q_{\nu_1 \nu_2 \cdots \nu_{s-1}}$  and the
analog of the Dirac quantization condition is \be \frac{1}{2 \pi
\hbar} Q_{\nu_1 \nu_2 \cdots \nu_{s-1}} P^{\nu_1 \nu_2 \cdots
\nu_{s-1}} \in Z . \ee

The case previously considered has dealt with black holes that may
possess both electric and magnetic $s = 1$ charges and the
underlying theory is the Einstein-Maxwell theory. It would be
natural to attempt an extension of the discussion to higher spins.
However, with our present stage of knowledge, this is only
possible for $s=2$. This is because we do not know how to couple
gravity to higher spin fields and sources.

{}For $s=2$, the theory exists and it is just the Einstein theory
where we know the analog of a black hole with both electric and
magnetic sources, which is the Taub-NUT space
\cite{Newman:1963yy,Misner1}. We also know that an ordinary
(``electric") test mass moves along a geodesic in a Taub-NUT
field.

With these two elements, the total angular momentum of the system
formed by a test electric mass at $r=c$ on the $z$-axis of a
magnetic Taub-NUT background was evaluated in \cite{BCHP}.  If,
following the customary notation that will be made explicit below
in (\ref{metric2}), the magnetic mass is denoted  by $N$ and the
test electric mass by $m$, the angular momentum is given by \be
J_z = 2 N m \label{Jz=2Nm} \ee and it is again independent of the
separation of the electric and magnetic masses.

Knowing (\ref{Jz=2Nm}) and using the insight provided by the
previous analysis, we will limit ourselves to describing, without
making any attempt to prove, what one expects to happen as the
mass is lowered on to the magnetic hole from a large distance.

When the mass is very far, the black hole is not rotating and at
distances $r_+ \leq r << c$,  the line element is that of a purely
magnetic Taub-NUT space, namely \beq ds^2 = - V(r)[ dt -
2N\cos\theta \, d\phi ]^2 + V(r)^{-1} dr^2
\nonumber \\
+(r^2+N^2)(d\theta^2 + \sin^2 \theta\, d\phi^2) \,,\hspace{1cm}
\label{metric1}\eeq with \be V(r) = 1 - \frac{2N^2 }{(r^2 + N^2)}
= \frac{r^2  -N^2}{r^2 + N^2}  \ee

At distances well beyond the electric mass, $ r >> c$, the metric
will asymptotically coincide with the leading approximation for
large $r$ of a Kerr-Taub-NUT space \cite{DemNew} with electric
mass $m$, magnetic mass $N$ and angular momentum $J_z = -ma = 2 N
m$, \beq ds^2 &=& -\frac{1}{\Sigma} \left(\Delta - a^2 \sin^2
\theta \right)dt^2 \nonumber \\ && + \frac{2}{\Sigma} \left[\Delta
\chi -
a (\Sigma + a \chi ) \sin^2 \theta \right] dt d \phi \nonumber \\
&& + \frac{1}{\Sigma} \left[ (\Sigma + a \chi)^2 \sin^2 \theta -
\chi^2 \Delta \right] d\phi^2 \nonumber \\
&& + \frac{\Sigma}{\Delta} dr^2 + \Sigma d \theta^2 .
\label{metric2} \eeq Here, we have set \beq && \Sigma = r^2 + (N - a \cos \theta)^2\\
&& \Delta = r^2 - 2 m r - N^2 + a^2 \\ && \chi = a \sin^2 \theta +
2 N \cos \theta \eeq

As the mass gets lowered, the outside Kerr-Taub-NUT approximation
gets better and better until it becomes exact when the mass
reaches the horizon.

In this case, just as in the electromagnetic case, one could say
that one got the black hole to turn more and more as the electric
mass approaches it.  This time, however, the ``transfer" of
angular momentum has not been from the electromagnetic field of
the source to the black hole, but rather, from the gravitational
field of the perturbation (which can be unambiguously separated
from the background) to the gravitational field of the hole.

\section{Conclusions}
\setcounter{equation}{0} \label{conclusions}

We have analyzed how the presence of an electric charge in its
exterior perturbs a magnetically charged, non-rotating black hole.
Because of the invariance of the equations under electromagnetic
duality, this situation is equivalent to placing a magnetic pole
outside an electrically charged black hole.  At large distances,
the geometry is that of a magnetic Kerr-Newman black hole.  When
the charge approaches the horizon and crosses it, the ``hair" is
lost and the exterior geometry becomes exactly that of a
Kerr-Newman hole with electric and magnetic charges and with total
angular momentum given by the standard value for a charged
monopole pair. Thus, in accordance with the ``no-hair theorem",
once the charge is captured by the black hole, the angular
momentum associated with the charge monopole system, looses all
traces of its exotic origin and it is perceived from the outside
as common rotation.

We have argued that a similar analysis performed on Taub-NUT space
should give the same result, namely, if one holds an ordinary mass
outside of the horizon of a Taub-NUT space with only magnetic
mass, the system, as seen from large distances, is endowed with an
angular momentum proportional to the product of the two kinds of
masses.  When the ordinary, electric, mass reaches the horizon,
the exterior metric becomes that of a rotating Taub-NUT space.
This rotating space (Kerr-Taub-NUT metric) is a solution of the
vacuum Einstein equations different from ordinary Taub-NUT space,
which, in spite of having both electric and magnetic mass, does
not possess angular momentum.

It is quite remarkable that one may set a black hole in rotation
by throwing into it radially a magnetic pole.  One may even
obtain, through successive applications of this process, a
rotating black hole which is neutral both electrically and
magnetically. Indeed, suppose that one starts with a Schwarzschild
hole and consider the following chain of four successive
processes:
\begin{enumerate} \item Throw in radially a charge $+q$ through
the North pole.  One then gets a Reissner-Nordstrom black hole
(electrically charged, non-rotating). \item Throw in now radially
a magnetic charge $+g$ through the North pole. One then gets a
Kerr-Newman black hole (electric charge $q$, magnetic charge $g$,
angular momentum $J_z = -\frac{gq}{4 \pi}$). \item Next, throw in
radially  through the South pole a magnetic charge $-g$.  One then
gets a Kerr-Newman black hole with no magnetic charge but rotating
twice as fast (electric charge $q$, magnetic charge zero, angular
momentum $J_z = -\frac{gq}{2 \pi}$). \item Finally, throw in
radially again through the South pole an electric charge $-q$. One
ends up with a Kerr black hole with vanishing total electric
charge, vanishing total magnetic charge and angular momentum $J_z
= -\frac{gq}{2 \pi}$. \end{enumerate} After this sequence of
processes is completed, it is impossible to tell that the Kerr
black hole that has been formed had anything to do with electric
or magnetic monopoles.  However, their existence was necessary to
set the black hole in rotation in this manner.

It is perhaps not totally inconceivable to imagine that our
universe is such that, at least part of the rotation of some of
the black holes that we have observed, might come from their
hiding the magnetic poles that we have not yet observed.

\section*{Note Added} After this work was put on the archive,
we learned of the interesting paper \cite{GarRey} where the
angular momentum of an electric charge at rest in the field of a
magnetic black hole is computed.  Our work is complementary to
this insightful article in that we consider in more detail the
dynamical transfer of angular momentum to the magnetic black hole
as the electric charge falls. We take a boundary condition
different from the one imposed in \cite{GarRey}, namely, we demand
that the hole is non rotating when the charge is very far from it.
On the other hand in \cite{GarRey} it is imposed that the spin of
the hole is zero when the charge is at a given distance $b$.  This
means that when the charge is at infinity, the hole is rotating in
such a way  that the sum of its initial angular momentum and the
angular momentum it receives when the charge gets from infinity to
$b$ exactly vanishes. For this reason the total angular momentum
depends on $b$ in their case while it does not in ours.  We also
check the disappearance of the associated hair when the black hole
forms by verifying the match of the relevant surface integrals on
the horizon of the Kerr-Newman black hole before and after the
electric charge has plunged in. We thank the authors of
\cite{GarRey} for calling our attention to their work.

Another paper that was brought to our attention after this work
was posted is  \cite{Friedman:1978zg}.   These authors discuss the
angular momentum of a test magnetic charge and a test electric
charge in a curved, axisymmetric (horizon-free) background
spacetime and verify that it is equal to the product $qg$,
independently of the curvature.  This is, in fact, a direct
consequence of our equation
(\ref{integrals forJz}) with $r_+ = 0$.
%The authors of
%\cite{Friedman:1978zg} do not investigate the problem of transfer of
%angular momentum to a charged black hole when a magnetic pole falls in and the associated
%disappearance of hair.

\vspace{0.5truecm}

\acknowledgments

This article was finished on board of the schooner ``Raquel".  We
thank the skipper and the crew for their generous hospitality and
support.  We also thank Rub\'en Portugues for key discussions in the
early stages of this research.  This work was funded by an
institutional grant to CECS of the Millennium Science Initiative,
Chile and also benefits from the generous support to CECS by
Empresas CMPC. The work of MH is partially supported  by IISN -
Belgium (convention 4.4505.86), by the ``Interuniversity Attraction
Poles Programme -- Belgian Science Policy" and by the European
Commission programme MRTN-CT-2004-005104, in which he is associated
to V.U. Brussel.

\vspace{1truecm}
\newpage

\begin{appendix}
\section{Establishing (\ref{spinvalue}) and (\ref{angularFieldvalue})}
\label{appendixA}\setcounter{equation}{0}

\subsection{Evaluation of (\ref{spinvalue})}
To compute the integral (\ref{spinvalue}), we first evaluate the
extrinsic curvature component \be K_{\phi r} = \frac{1}{2N} \left(
N_{\phi \vert r} + N_{r \vert \phi} \right) \ee of the slices t=
const of the metric (\ref{KerrNewman}).  One has \be g_{\phi r} =
- \frac{2 M a}{r} \sin^2 \theta + \frac{g^2 a}{r^2} \sin^2 \theta
\ee and hence \be N_{\phi \vert r} + N_{r \vert \phi} = \frac{6 M
a}{r^2} \sin^2 \theta - \frac{4 g^2 a}{r^3} \sin^2 \theta \ee  The
lapse is equal to $$N = \left(1 - \frac{2 M}{r} +
\frac{g^2}{r^2}\right)^{1/2}$$ (to first order in $q$) and hence
\be K_{\phi r} = \frac{\frac{6 M a}{r^2} \sin^2 \theta - \frac{4
g^2 a}{r^3} \sin^2 \theta }{2 \left(1 - \frac{2 M}{r} +
\frac{g^2}{r^2}\right)^{1/2}} \ee  The momentum component
$\pi_\phi^{\, r}$ is related to $K_{\phi r}$ by \be \pi_\phi^{\,
r} = - \frac{1}{16 \pi} K_{\phi r} g^{rr} \sqrt{g} \ee which leads
to \be - 32 \pi \, \pi_\phi^{\, r} =  6 M a \sin^3 \theta -
\frac{4 g^2 a}{r} \sin^3 \theta .\ee  Integration over the angles
gives then the announced result at $r = r_+$, \be 2 \int_{S(r_+)}
\pi_\phi^{\, r} d \theta d \phi = - M a + \frac{2}{3} \frac{g^2
a}{r_+} . \ee

\subsection{Direct evaluation of (\ref{angularFieldvalue})}
We evaluate directly in this appendix the difference $\Phi(\infty)
- \Phi(r_+)$ when the electric charge is near the horizon, $c =
r_+ + \epsilon$.  To that end, we observe that $\Phi(r)$ is a
continuous function at $r = c$ (the discontinuity in the Heaviside
function is compensated by the same discontinuity in the integral
of $\mathcal{E}^r$). This statement was proven for flat space in
section {\bf \ref{Flat}} (recall the discussion following
(\ref{Flux0})).  This continuity stays valid when the test charge
is placed on the curved background of the magnetic pole because at
the location of the electric charge, that background can be
obtained by a smooth deformation of flat space.  Therefore, we can
compute equivalently $\Phi(\infty) - \Phi(r_+)$ when the electric
charge has just plunged into the black hole.  In that case,
$\Phi(\infty) - \Phi(r_+)$ reduces to \be \Phi(\infty) - \Phi(r_+)
= \frac{g}{2}\int_0^\pi d \theta \cos \theta \mathcal{E}^r(r_+,
\theta) \label{AABB}\ee where the electric field is that of a
Kerr-Newman black hole with electric charge $q$, magnetic charge
$g$ and angular momentum $qg$.

The radial component $\mathcal{E}^r$ of the electric field at
$r_+$ has two pieces: one, $\mathcal{E}^r_1$, which comes from the
electric charge and another, $\mathcal{E}^r_2$, which comes from
the rotation of the magnetic hole. To first order in $q$, they are
given by \beq \mathcal{E}^r_1 &=& q \sin \theta \\ \mathcal{E}^r_2
&=& - \frac{2 g a}{r_+} \cos \theta \sin \theta . \eeq Only $
\mathcal{E}^r_2 $ contributes to the integral (\ref{AABB}),  \beq
\Phi(\infty) - \Phi(r_+) &=& -\frac{g^2
a}{r_+}\int_0^\pi d \theta \cos^2 \theta \sin \theta \nonumber \\
&=& - \frac{2}{3} \frac{g^2 a}{r_+}, \label{final}\eeq which is the
announced formula (\ref{angularFieldvalue}).

Strictly speaking, once (\ref{spinvalue}) is proven, it is not
necessary to establish (\ref{angularFieldvalue}) directly because it
follows from (\ref{integrals forJz})  and (\ref{valueofJ}) that \be
\Phi(\infty) - \Phi(r_+) = J_z - 2 \int_{S(r_+)} \pi_\phi^{\, r} d
\theta d \phi\ee and $J_z = -Ma$.  We have included nevertheless
this derivation because we believe that it provides additional
insight on the mechanism through which the capture of a magnetic
pole sets a black hole in rotation.

\end{appendix}

\end{document}